\documentclass[10pt]{article} 
\usepackage[english]{babel}
\usepackage{graphicx}
\usepackage{amsmath}
\usepackage{amssymb}
\usepackage{mathtools}
\usepackage{braket}
\usepackage{setspace}
\usepackage[utf8]{inputenc}
\usepackage[makeroom]{cancel}
\usepackage{enumerate}
\usepackage{cite}

\usepackage[top=2.5cm, bottom=2.5cm, left=2.5cm, right=2.5cm]{geometry}

\usepackage{color}
\usepackage[colorlinks=true,
            linkcolor=red,
            urlcolor=blue,
            citecolor=lightblue]{hyperref}
\definecolor{darkblue}{rgb}{0.0, 0.0, 0.5}
\definecolor{darkblue}{RGB}{0,0,80}
\definecolor{lightblue}{RGB}{50,120,150}
\definecolor{darkgreen}{RGB}{0,80,0}
\definecolor{darkred}{RGB}{80,0,0}
\definecolor{amber}{rgb}{1.0, 0.75, 0.0}
\definecolor{arsenic}{RGB}{50,30,90}
\definecolor{ao(english)}{rgb}{0.0, 0.5, 0.0}

\usepackage{xparse}
\usepackage{cleveref}
 
\begin{document}
 
\title{\textsc{Lithium enrichment threatens to curb fusion deployment}}

\author{S. H. Ward\textsuperscript{1}, R. J. Pearson\textsuperscript{2}, T. Scott\textsuperscript{2}, N. J. Lopes Cardozo\textsuperscript{1}\\ \\
1. Eindhoven University of Technology, Eindhoven, Netherlands\\
2. School of Physics, University of Bristol, Bristol, United Kingdom\\}

\renewcommand{\thefootnote}{\arabic{footnote}}
\maketitle 

\begin{abstract}
\noindent  
The impact of lithium isotopic enrichment on the global deployment of nuclear fusion energy is analysed. Lithium -- the $^6$Li isotope in particular -- is essentially one of two elemental fuels required by fusion reactors for tritium breeding. Whilst variable \textit{consumption} of lithium is low enough to present negligible cost, it is instead the large stored inventory \textit{volume} (50-100 tonnes) and its \textit{required enrichment} that compound to significantly drive capital costs. These costs are driven by the inefficiency of the tritium breeding process, making this challenge fundamental to almost all fusion power plant concepts. Financing would further compound these effects, making lithium fusion fuels more akin to an upfront capital expenditure than operational expenditure. 

Other potential barriers to fusion deployment created by lithium are also discussed: enrichment technologies of today are shown to be too expensive, not scalable, and environmentally risky, and highly enriched $^6$Li is a controlled substance. Mitigating actions include: developing alternative enrichment technologies that are affordable, scalable, and do not rely on mercury; incorporating lithium enrichment as an explicit cost driver in reactor design processes, producing more compact reactors with smaller lithium inventories; establishing distinct enrichment levels to enable supply chain monitoring for misuse; and the most radical solution: breeding blankets that use natural, unenriched lithium. These actions may impact tritium breeding capabilities, which calls for an urgent re-assessment of the tritium breeding paradigm. Whatever solution is sought, lithium supply is a mission-critical issue that needs urgently addressing.
\end{abstract}

\section{Context \& scale}\label{contextscale}

The commercial deployment of fusion energy relies on the availability and enrichment of $^6$Li for tritium breeding, essential for fuelling fusion reactions. With a new fusion industry emerging and demonstrators aiming to operate in the next decade, $^6$Li availability poses a significant bottleneck. Historical mercury-based enrichment methods are unsustainable and may not scale. There is an urgent need for innovative, scalable, and environmentally safe lithium enrichment technologies, but cost, as well as geopolitical and non-proliferation implications, demand careful consideration. A collaborative, multidisciplinary approach is essential.

We examine alternative pathways utilising natural or lower-enriched lithium, advocating for a broader reassessment of breeding blanket design philosophy. $^6$Li is an under-explored but key challenge in achieving the global deployment of fusion, a technology with significant potential for providing sustainable energy for humankind.

\section{Introduction}
A fusion industry requires materials for the construction and operation of power plants. Whether these materials become limiting factors has been addressed in the literature (e.g. \cite{Pearson2022}\cite{Schleisner2001}\cite{nicholas2021}). So far, relatively few materials are identified as critical. Lithium is one of them. 

Lithium is effectively one of the primary fuels of the fusion reactor; In the internal fuel cycle, neutrons released by fusion events react with lithium to produce tritium, which is then separated to feed the deuterium-tritium fusion reaction. For this cycle to perpetuate, at least one triton must be bred per fusion neutron released - a tritium breeding ratio (TBR) of unity. In reality, the TBR must be greater (by some 20\%) to account for tritons and neutrons lost during this process. Lithium is naturally composed of the stable isotopes $^6$Li and $^7$Li in ratios of approximately 7.5 \% and 92.5\%, respectively. The neutron capture reaction with $^6$Li is exothermic, releasing 4.8 MeV (up to tens of percent of reactor power output), and has a cross-section for thermal neutrons of $\sim$ 1000 barn at room temperature. The reaction with $^7$Li has a threshold energy of 2.5 MeV and reaches the much smaller cross-section of $\sim$ 0.3 barn in the relevant energy range 5 -- 14 MeV, but has the advantage of producing a secondary neutron in addition to a tritium. Both isotopes are therefore useful in the tritium breeding blanket. However, as neutrons slow during their passage through the breeding zone, the largest contribution to the tritium production comes from the $^6$Li isotope. Thus to maximise TBR, blankets are often predicted to require $^6$Li enrichment levels of $60 - 90 \%$ (or potentially as low as 30\% for solid-based breeders with neutron multipliers).

In this paper we broaden the current discussion of lithium availability. In the economic analysis, we compare the cost of capital and lithium enrichment to consumption. We also examine lithium in a wider context to discuss its impact on the fusion value proposition. With energy independence often regarded as a key asset, fusion energy must not depend on monopolised materials, and estimates of lithium availability must account for an increasingly varied geopolitical domain. Likewise, supply chains must be considered wholly when assessing fusion energy's credentials as a clean and safe source of power -- this includes environmental and climate impacts of lithium mining and processing. This also applies to scalability: if the processing of lithium requires scarce substances, then lithium itself must be considered scarce even if its unprocessed reserves are large. In all of these considerations we distinguish the short term (is there enough to build a few DEMO reactors?) from large scale deployment, which calls for more than a hundred fusion power plants constructed per year. We base our analysis on literature, especially the recent overview papers on the enrichment of lithium \cite{Giegerich2019}\cite{Ault2012}\cite{Badea2023}\cite{Murali2021}\cite{Wang2022a}\cite{Wang2022b}\cite{Xiao2017}.

The \textit{consumption} of lithium, at about 100 kg per year for a 1 GWth (effective) power plant, is small. Terrestrial reserves of lithium are not considered a limiting factor for the foreseeable future, nor would lithium consumption incur significant costs.
The lithium consumption of a 1 GWth fusion power plant that generates electricity corresponds to roughly $2.10^{7}$ kWh per kg of lithium (assuming a $\sim20\%$ overall plant efficiency \cite{Mulder2021}). At a cost of electricity (CoE) of \$0.1/kWh, the generated revenue is $\sim$ \$2M per kg of lithium. Hence, fusion could pay \$200k/kg and the variable cost of fuel would still be a 10\% contribution to the CoE. With the present market price of purified metallic lithium, hundreds of \$/kg, with the cost of the lithium component roughly five times larger, the cost of lithium would appear to be insignificant. This estimation neglects the fact that the financing of the required reactor inventory of lithium will cost two orders of magnitude more than the consumption. The allowable cost of purified lithium would therefore be of the order \$1-10k/kg -- at most two orders of magnitude above present market price. This raises the question of future lithium price volatility given the potentially stable costs for enrichment and purification.\\

It has been suggested \cite{Bradshaw2011} that lithium consumption by the battery industry could leave none remaining for fusion. Certainly, demand for lithium is expected to strongly increase in the coming decades, to more than a million tonnes per year by 2040. However, it is unlikely that increased demand drives lithium prices too high for fusion. Consider the acceptable price of a car: the cost of (natural) lithium cannot rise above a ceiling of roughly \$200/kg, corresponding to 10 k\$ for the lithium in a battery pack containing 50 kg of lithium. Therefore, the value generated by burning lithium in a fusion plant far outweighs the highest cost that the car industry is willing to pay; fusion firms can easily out-bid battery manufacturers -- to the point of purchasing premium electric vehicles purely for the lithium in their batteries. Instead, external demand for lithium generates more value for fusion than costs; prior extraction and processing of lithium by other industries will reduce the time and risk to making lithium available -- two things a fusion industry is far more sensitive to than price. Hence the electric vehicle industry should be reframed as a pump primer, not competitor. Even if fusion power were fully deployed ($\sim$ ten thousand plants), the total fusion lithium inventory would still be a small fraction of the lithium within one billion electric vehicles.

In summary so far, neither abundance nor cost would produce major bottlenecks if fusion power plants were to use natural lithium. And in as far as fusion would be a much smaller user of lithium than competing industries, its production would contribute relatively modestly to any environmental footprint.

However, most fusion power plant concepts don't use natural, but highly enriched lithium, in which the relative content of the $^6$Li isotope is enriched from the natural 7.5\% to typically more than 50\% -- up to 90\% has been suggested for some concepts based on the use of liquid lead-lithium (ref\cite{Giegerich2019} and references therein). Hence, now the \textit{enrichment process} should be considered as a potentially limiting factor. Today, these processes are not economically viable, scalable or environmentally responsible. 

\begin{figure}
    \centering
    \includegraphics[width=1\linewidth]{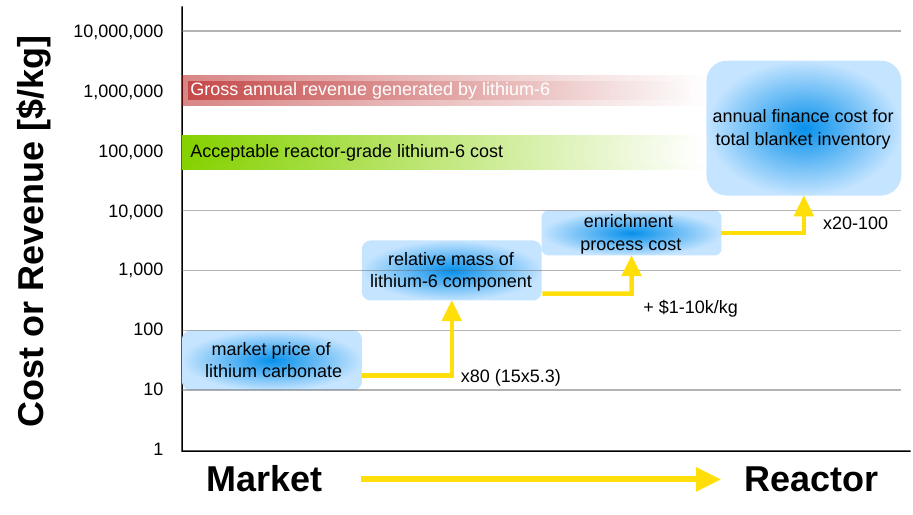}
    \caption{This graph illustrates the costs and benefits arising from lithium fuel. While the cost of the annual consumption of lithium is orders of magnitude smaller than the generated revenues, the cost of enriched lithium -- specifically financing the procurement of 1 tonne of it for every kg of annual use -- is in a range where it could jeopardise the economic viability of fusion power. The interest rate herein varies between 2-10\%.}
    \label{fig:costbuildup}
\end{figure}

A compounding factor is the quantity of $^6$Li needed in a fusion power plant. The designs for blankets in demonstration-scale devices (DEMO) contain roughly 50 tonnes of $^6$Li. For power plant blankets that must be replaced and reprocessed every 5 years, and contain lithium purchased and processed prior to assembly, approximately 100 tonnes of $^6$Li are required per GWth-scale reactor. Annual consumption of $^6$Li is therefore only one thousandth of the site inventory, making the system extremely capital inefficient. So interest from financing upfront inventories dominates annual costs arising from $^6$Li usage. The price of $^6$Li itself is made up of two components: the relative abundance of $^6$Li; and the cost of enacting the enrichment (and any purification), which is uncertain but reviewed herein. Figure \ref{fig:costbuildup} illustrates this cost breakdown.

A single EU-type DEMO-scale device would demand far more enriched lithium than currently supplied by the global enrichment capacity - which is near negligible \cite{Pearson2022}. This does not account for the other DEMOs currently planned (for example the U.K. STEP device). For a first generation of fusion power plants, projected a decade after these DEMOs, the shortfall is an order of magnitude larger. Lithium enrichment is therefore an urgent issue, on the critical path of fusion deployment.


\section{Three factors that make lithium critical: scale, enrichment, \& upfront investment}

\subsection{Scale, scalability and the required rate of upscaling; the ten-thousand plants perspective}\label{Scale}
We focus on two phases of fusion deployment. After successful demonstrators may confirm fusion as potential energy source, a growth phase would follow, during which the global fusion industry must sufficiently expand to meet demand for future power plants (FPPs). Approximately ten-thousand GWth-scale plants must operate to supply 20\% of current, global primary energy demand \cite{GlobalPrimaryEnergyDemand}. Sustaining this number requires an industry to produce 200 FPPs per year, assuming a lifetime of 50 years. At an overnight investment of several G\$ each (see e.g. \cite{Entler2018}\cite{Griffiths2022}), that defines an industry with a total turnover of order 1 trillion USD per year. In comparison, the nuclear fission industry today has capacity to produce $\sim 10$ new power plants annually.
Nevertheless, this industry will not appear overnight. Building it from a few demonstration reactors, over roughly 10 years, to the target of 200 FPPs per year involves 7 - 8 doublings. Typically, such development appears exponential, with a doubling time of $\sim$ years \cite{Kramer2009}\cite{LopesCardozo2016}\cite{LopesCardozo2019}. As such, about two decades are required to scale industrial capacity. 
Focussing on the early phase of development, after the commissioning of successful demonstrators, we may suppose that a first generation of some 10 commercial fusion power plants could operate before 2050.
This scenario would align with the timelines of the more ambitious national strategies, for example: the 2018 European road map \cite{EUROfusion2018}, and its accelerated update since; the U.S.fusion strategy \cite{El-Guebaly2018}; the U.K. strategy \cite{UKStrategy2022} and its streamlined regulation and the new strategy of China.\\

Figure \ref{fig:upscaling} illustrates this deployment path, where installed fusion power increases ten-fold every decade, starting with ten Gen1 plants in 2050, following DEMO projects that operate by 2040. To enable this scenario, in excess of 100 tonnes of enriched $^6$Li must be available 10 years from now, as DEMO reactors begin construction. The corresponding production rate of enriched $^6$Li compares to that of the U.S. in the 1950s for the production of nuclear weapons -- the largest in history. For the mercury-based enrichment methods of today, this would require 10 kt of mercury -- roughly 25 times today's annual world production. This enrichment capacity would then need to increase tenfold every decade, meaning any recycled mercury could offset only approximately 10\% of demand at any one time. 

In conclusion, analysing the availability of lithium requires estimating: the scale of the required fusion deployment, the scalability of all processes involved in the supply chains, and the speed at which up-scaling can realistically be achieved. \\

\subsection{Enrichment}\label{enrichment}
\subsubsection{Mercury-based enrichment}
Lithium enrichment at industrial scale has been and is being done exclusively for the production of nuclear weapons (see e.g. \cite{USGAO2013}). The COLEX process was selected in the U.S. after the most efficient enrichment methods were reviewed and explored in the 1950s. COLEX utilises the difference between the $^6$Li and $^7$Li isotopes' chemical affinity to mercury. This same process is presently used for lithium enrichment in China, North Korea \cite{Albright2017} and Russia. Although publicly available information about these enrichment programmes is scarce, an estimate of the amount of mercury needed for an annual production of 1 tonne of highly enriched $^6$Li can be gleaned from the information about the U.S. programme at Oak Ridge. Between 1953 and 1963, this plant produced 442 tonnes of 6-lithium hydroxide -- about 100 tonnes of $^6$Li, for which it used 11 thousand tonnes of mercury \cite{Kramer2018}\cite{Giegerich2019}. This equates to 0.3 - 1 kt of mercury for the enrichment of 1 tonne of $^6$Li per year. Scarce information about the North Korean programme \cite{Albright2017} would be consistent with that estimation. 
In a recent review paper, Giegerich and Day \cite{Giegerich2019} compared all available enrichment technologies and concluded that the COLEX process, or rather a cleaner, more modernised adaptation -- the 'improved column-based mercury amalgam exchange' (ICOMAX) -- still offers the most viable route to industrial scale $^6$Li production. Other reviews, and recent industry initiatives, such as the UKAEA's Fusion Industry Programme, consider techniques based on lasers, chemical crown-ethers, and even microbes, but none of these have been publicly demonstrated at scale. \\
 
\noindent For a fusion power plant blanket containing 100 tonnes of $^6$Li, an enrichment plant with some 3 - 10 kt of mercury, i.e. the size of the 1950s Oak Ridge Y-12 facility, would be required during a period of 10 years. These are large numbers that call for some critical reflections.\\

\setlength{\leftskip}{1cm}

\noindent\textbf{First}, enrichment of $^6$Li using the mercury process is expensive: prior to its banning from open markets, mercury costs neared \$30k/tonne. At 7\%, the annual cost of financing this investment exceeds \$2k per kg of enriched $^6$Li. This is excluding the construction, running and decommissioning of the enrichment plant. In addition, unless the low volumes of depleted lithium can be sold at high price, the cost of the raw $^6$Li component of natural lithium is 15 times that of pure, natural lithium (roughly 80 times the cost of raw lithium carbonate) -- putting the price floor of $^6$Li near \$1-2000/kg. However, a price floor of several k\$/kg places $^6$Li in the critical cost bracket where it significantly contributes to the cost of fusion powers. To put this in perspective: today $^6$Li is sold at about \$50k/kg, in small quantities only, mostly for medical purposes. 
\\

\begin{figure}
    \centering
    \includegraphics[width=0.85\linewidth]{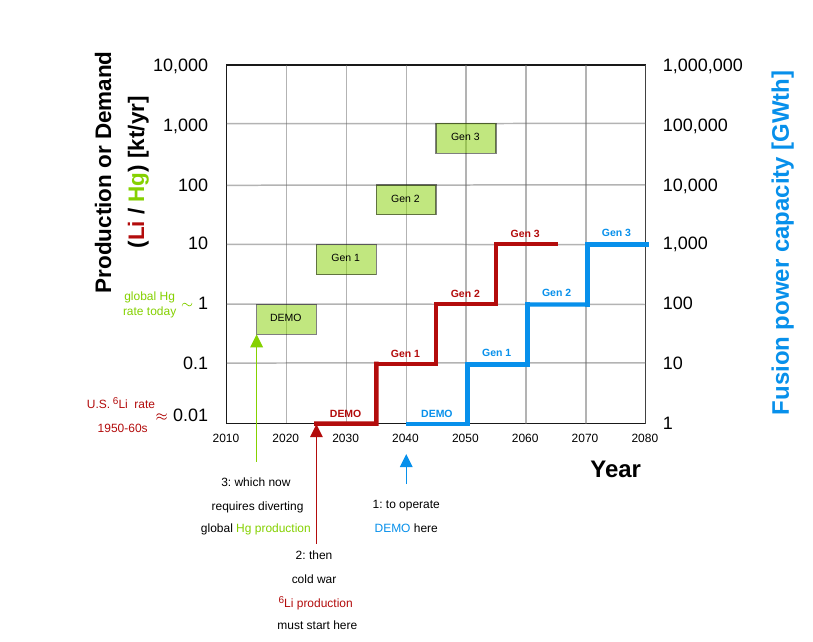}
    \caption{To enable the deployment of fusion power, based on tritium breeding blankets containing enriched $^6$Li, the enrichment capacity should already now compare to cold war production at the Y-12 facility in Oak Ridge, growing ten-fold every decade. If this enrichment were to use the mercury-based process, the required mercury production would soon outstrip the annual global production. The bandwidth shown in mercury production rate reflects an ambiguity in the literature.}
    \label{fig:upscaling}
\end{figure}

\noindent\textbf{Second}, mercury-based enrichment is unlikely to scale: a single DEMO plant requires a large quantity of mercury -- 3-10 kt, compared with current global annual production of 2 kt. Scaling up to full deployment would require another factor of 200 increase of the mercury production, making it 3 orders of magnitude larger than the present global production. Hence present mercury-based enrichment technologies are not scalable to the required volume. They fall short by at least 2 orders of magnitude and therefore cannot support large-scale deployment of fusion power. Figure \ref{fig:upscaling} illustrates the fundamental issues with the scalability of lithium enrichment.\\

\noindent\textbf{Third}, the enrichment process is not environmentally responsible; usage of the COLEX process in 1950s and 1960s America had a large detrimental environmental impact\cite{Brooks2011}. During that period, about 10\% of the mercury was lost, leading to pollution severe enough to lead to the United States banning COLEX (now also covered by the United Nations Minamata Convention, which aims to ban the new production and trade of mercury and non-military products that contain it). Giegerich and Day \cite{Giegerich2019} argue that, with modern technology, an environmentally responsible process can be devised (ICOMAX). Nevertheless, mining and handling of mercury would still be required -- activities that are already hazardous at present extraction rates \cite{UNEP2018}.\\

\noindent\textbf{Fourth}, mercury-based enrichment is geographically localised and geopolitically sensitive: China alone accounts for roughly 90\% of global mercury production\cite{sodeno2023projected} - though China also became a signatory of the Minimata Convention as of 2013. Therefore, mercury-based enrichment processes reduce the ability for fusion to improve energy independence.\\

\setlength{\leftskip}{0cm}
\noindent In summary, the mercury-based enrichment process is too expensive for commercial fusion power, environmentally irresponsible, prone to geopolitics and not scalable to the production of more than a few fusion power plants per decade. Nevertheless, it is currently the only proven industrial-scale enrichment method, and so there remains value in deploying intermediate, mercury-based processes -- if they can improve upon the aforementioned qualities.

\subsubsection{Alternative enrichment processes}
If mercury-based enrichment is not a feasible route, are there any other candidates? In a review of isotope separation options, Symons \cite{Symons1985} in 1985 already stated that a mercury-based process is not scalable, and expresses the hope that laser-based processes would become feasible in the 21st century. 
More recent literature reviewed the options \cite{Giegerich2019}\cite{Ault2012}\cite{Badea2023}. Ault et al \cite{Ault2012} focus on the complementary goal to purify $^7$Li for use in nuclear fission power plants. They reject the mercury-based process and give a detailed analysis of two known alternative technologies: the laser-based AVLIS (atomic vapor laser isotope separation) and a chemical process based on crown ethers. The latter is judged to be technically feasible at the required scale, but would require a large cost reduction to be economical. For the AVLIS technology they express concerns about the achievable enrichment. For both technologies, scaling up by orders of magnitude would be required. Badea et al \cite{Badea2023} compare different groups of separation methods, looking at the separation factor, the complexity of the method and cost, as well as the environmental impact. The chemical exchange methods using crown ethers are found to be effective but expensive and environmentally risky, whereas the less costly and environmentally safer methods -- electrochemical exchange and displacement chromatography -- score low on separation factor. They conclude that the crown ether method is currently the most technically viable, despite its high cost and environmental impact.
Giegerich and Day \cite{Giegerich2019} also compare the available separation technologies and, using a rubric, conclude that the mercury-based process is the best option. They propose to develop the improved mercury-based ICOMAX process to provide the enriched $^6$Li for EU-DEMO and describe a development programme that would have enrichment beginning by 2040 and completing supply of a DEMO blanket by 2045. Note that this plan assumes DEMO commissioning starts in 2050 -- much later than more recent road maps. Finally, an entirely different approach is advocated by \cite{Diaz-Alejo2021}, who explore the use of micro-algae to separate out the $^6$Li isotope. Whilst such nuclear biotechnology is in its infancy, it would appear to score well on scalability, affordability and environmental impact.

In summary, these studies confirm that only the COLEX process is presently able to provide enriched $^6$Li at scale, and all agree that an alternative is necessary. Giegerich and Day seek the solution in an improved mercury-based process. However, as we have shown above, while this may provide enough enriched lithium for a DEMO reactor, albeit via problematic supply chains, it is not capable of scaling to meet the demands of a fusion power industry. Ault et al \cite{Ault2012} and Badea et al \cite{Badea2023} favour the crown ether method (which comes out as 2nd best in \cite{Giegerich2019}), but all authors note that it is very costly, not eco-friendly, and not available at significant scale. In view of our cost and scaling arguments, this method would also face significant issues when the basis of a fusion industry. Finally, neither of these alternatives to COLEX are available at a relevant industrial scale today, which places lithium enrichment firmly on the critical path. The UKAEA open call to industry to propose and develop lithium enrichment technologies, which resulted in six companies receiving funding for such development, is one of the first recent acknowledgements of this.

\subsection{CAPEX: Possession costs more than consumption}\label{possession} 

Enriched lithium is a significant upfront cost component of a fusion power plant. The blanket contains typically 2-3 orders of magnitude more lithium than the annual consumption (of order 10 plant lifetimes of supply). Two blankets are effectively needed for a power plant to operate continuously: one for operation and at least one in standby (possibly undergoing reprocessing which must happen prior to replacement). And with blankets requiring decadal timescales for construction, and operators potentially favouring shorter, high-frequency maintenance cycles \cite{schwartz2024valuing}, all blankets should be prepared prior to operation. Phrased differently, enriched lithium must be procured well in advance of operation, and after blanket replacement it remains in possession for reprocessing and re-use. Hence blankets are productive -- at most -- only 50\% of the time; any increase in blanket failure rates (from 0\%) or reduction in overall plant capacity (from 100\%) will reduce this productivity factor and decrease capital efficiency.

This results in 100 tonnes of $^6$Li per reactor per GWth generated power.
The annual consumption -- about 100 kg per year per GWth -- is about 0.1\% of the total inventory. Hence at an interest rate of 7\%, the cost of capital is 70x the cost of the consumption. This will increase after accounting for availability and capacity factor: if the plant is not running it does not consume lithium nor does it generate revenues, but debt servicing must continue.

Alternatively, one may consider the overnight capital investment of a fusion power plant. To be comparable to fission, investment costs must lie near \$5 - 10 per We effective power. That corresponds to an investment of some 2 billion USD for a 1 GWth reactor. For example Entler et al \cite{Entler2018} use a figure of \$8.5B for overnight investment, for a 4.1 GWth fusion power plant, to which they ascribe 0.95 GWe output power at 75\% availability. At a cost level of \$5k per kg, 100 tonnes of $^6$Li would contribute 25\% to the total overnight cost of the power plant. This figure could increase if fusion plant costs follow a learning curve. As a side note, the cost of enriched lithium is not explicitly included in the cost estimates of \cite{Entler2018}, and neither is the blanket reprocessing cost.

	
\section{Is lithium enrichment an absolute necessity?}
Blanket concepts that operate with natural lithium would avoid the aforementioned issues. Such concepts do exist but must demonstrate a sufficient tritium breeding ratio (TBR) is achievable. In the analysis of \cite{Pereslavtsev2016}, in the case of a helium-cooled pebble bed DEMO divertor, the assumed 60\% enrichment adds about 25\% to the TBR. This raises the question: is it necessary to increase the TBR by 25\%? In a forthcoming paper, we discuss ideas that can reduce the required TBR, thus facilitating natural lithium-based breeder concepts in toroidal devices \cite{TUE_tritium_lean_startup}.

There is limited literature on breeders based on natural lithium, for example \cite{Nishikawa1989} and \cite{Fierro2020}, the latter of which concerns a blanket concept for inertial fusion. Generally, blankets in pulsed and inertial-type devices could have superior performance to those in toroidal devices due to a higher solid angle coverage; breeding material in toroidal devices must be supplanted with other components not required in pulsed schemes, for example divertors and central solenoids. Another factor that influences TBR and thus blanket design are the first wall materials in which neutrons slow down or are absorbed. The first wall thickness plays a role in influencing TBR, with every centimetre of steel first wall decreasing the TBR by some 7-10\% \cite{Pereslavtsev2016}\cite{Dai2021}. The same principles apply for nearby components, including the structural materials of the blanket and any plasma-facing tungsten armour. 

Reactors that could employ thinner walls, or avoid first walls altogether (for example, present designs proposed by First Light Fusion, General Fusion, Zap Energy, and Renaissance Fusion), have an advantage when maximising TBRs using natural lithium. Though the viability of these concepts has not yet been demonstrated, their circumvention of lithium enrichment challenges should be considered valuable at the strategic level \cite{TUE_value_parallel}.

\section{Non-proliferation considerations}\label{nonproliferation}

High-purity $^6$Li will be required by fusion developers to validate breeder blanket concepts, before orders of magnitude more would need to spread globally to fuel a future industry. But procuring or trading such isotopically rare materials is deemed to pose nuclear proliferation risks today; such highly isotopically concentrated $^6$Li is already in use in nuclear weapons devices, and it is typically covered by export controls. Therefore, there is a trade-off between commercial market reach and the fuel cycle performance of individual reactors. Hence a question arises: what is the permissable concentration of $^6$Li in a commercial setting? 

The required level of regulatory oversight for fusion power plants has not been determined, and some imply that stringent controls or safeguarding are unnecessary. Further work is needed to fully understand and demonstrate the non-proliferation potential of fusion, but an instructive analogy can be drawn with fission fuel production, where the fissile isotope $^{235}$U is used in both nuclear weapons and nuclear fuel. In conventional light water reactors, uranium dioxide fuels have an isotopic enrichment of approximately 3.5\%, increasing to 20\% for HALEU fuels (high-assay low-enriched uranium). In contrast, the enrichment level of $^{235}$U in nuclear weapons exceeds 95\%. Consequently, it is evident to nuclear inspectors, based on sampling and mass spectrometry analysis, whether an isotopic enrichment facility is being used for civilian or military purposes.

While $^6$Li is not controlled in the same way as fissile isotopes of uranium, there are controls on its production, supply, and use due to its utility in nuclear weapons (both for tritium production and use in thermonuclear devices). For lithium as a fuel for fusion, could it be that there is an appropriate level of enrichment that enables a clear distinction between lithium modified for fuel applications versus weapons applications? It is possible that there needs to be a separation between what could be described as "fusion grade" and "weapons grade" $^6$Li. The fusion community should focus on identifying enrichment levels that allow for technically feasible (net positive) tritium breeding, yet which are demonstrably far from the high purity required for potential weapons applications. Here, the cutoff for what constitutes "high purity" must be determined, as lower enrichment levels for breeder blankets appear favourable, but there could be significant impact on designs depending on whether this is set at 20\% or 60\%. Some level of monitoring and inspection for the $^6$Li supply chain appears necessary, but such a distinct separation of two grades of $^6$Li could serve to not only address security concerns but also aid in public acceptance by demonstrating a conscious effort to avoid links with weapons-related isotopic purities.

In the case that the deployment of fusion becomes widespread, given that all lithium enriched above natural levels is currently export-controlled, it is worth considering whether the current regime could be revisited for fusion purposes. Enrichment below some threshold might warrant reduced controls, enabling broader international distribution, potentially limited to nations complying with the nuclear non-proliferation treaty. This raises the question of whether fusion can achieve the goal of being a globally deployable energy source if its rollout is restricted by country compliance, i.e. that abundant fusion energy is available only to countries adhering to a centralised, rules-based order. Further, it highlights that fusion still requires responsible end users, contrasting with the benign nature of other renewable energy sources. One possible solution is that the $^6$Li supply chain could be controlled by embedding enriched lithium in forms that are difficult to separate, such as lithium-lead mixtures, solid compound forms, and molten salts like FLiBe -- all leading breeder candidates for commercial fusion. This approach sets precedent in the "de minimis" rule, where security controls are reduced if the controlled substance is within a broader material mix and difficult to extract. Judging breeders this way, based on their chemical make-up, would have implications for blankets based on pure lithium.

Non-proliferation considerations could reduce the maximum permissable $^6$Li concentration for fusion device development. Maximising the cleanliness, safety, and coordination of a fusion industry requires transparently addressing the proliferation risks of $^6$Li usage. The public should understand that, while fusion has inherent safety features, mitigating further risks requires careful management of $^6$Li, amongst other nuclear materials. Close collaboration amongst fusion experts, security specialists, policymakers, and industry leaders is needed to develop a viable path forward.

\section{Conclusion}\label{conclusion}

Current lithium enrichment technologies do not allow for fusion power deployment at a meaningful scale or speed. Former production of $^6$Li by mercury-based enrichment is unlikely to scale, is expected to be too expensive, and introduces environmental and geopolitical risks. Other enrichment technologies exist but have not yet been proven at benchtop scale, let alone industrial scale. This issue is slowly drawing attention; the U.K. government has prioritised funding for innovative R\&D exploring new low-cost, scalable, and environmentally friendly solutions. However, even if successfully delivered at scale, any solutions resulting in a $^6$Li cost in the order of thousands of dollars per kg will significantly contribute to the cost of fusion energy.

To enable the first phase of fusion deployment, several thousands of kilotons of $^6$Li must be available -- an order of magnitude more than the flagship U.S. Y-12 facility produced in the 1950s and 1960s. Any new enrichment technology must therefore be pursued urgently to prevent a bottleneck for global deployment of commercial fusion power.

Designs with smaller $^6$Li inventories or smaller blankets relative to reactor size (large-scale devices) may mitigate the issue but will not remove the fundamental problem of scalability to full deployment. Further still, breeding blanket concepts based on the use of natural lithium may avoid the aforementioned issues, but potentially at a cost to the tritium breeding performance for a viable closed tritium fuel cycle. Whilst this cost may be worthwhile, future work should review the requirement for such a high tritium breeding ratio. If this requirement could be relaxed, and natural lithium employed, it would also mitigate any potential downside risks in a scenario where fusion power cannot be demonstrated in the medium--long term; in this case, a low-cost and scalable method for enriching sensitive material may be unwanted \cite{baum2014great}.

Proliferation concerns may also narrow the market for any fusion reactor which uses any level of enriched $^6$Li. However, identifying enrichment levels that balance technical feasibility with non-proliferation can facilitate public acceptance and security, which must be considered as integral to the process rather than an afterthought.

Despite this, the challenges outlined are not insurmountable and there are opportunities for new technological approaches. Instead, these challenges act to reformulate the value proposition for different nuclear fusion reactor approaches based on their sub-technologies. Innovative reactor designs, improving enrichment technologies, and ensuring robust non-proliferation measures must be considered urgently and in parallel. Collaboration between experts, security specialists, policymakers, and industry leaders is essential to navigate these challenges to avoid $^6$Li enrichment becoming a critical bottleneck for the deployment of fusion, and to realise the full potential of fusion energy as a clean, safe and secure power source for humankind.

\subsection{Acknowledgements}\label{acknowledgements}

The authors would like to thank Norbert Wegrzynowski (University of Bristol) for their comments on mercury supply, Thomas Giegerich (KIT, Germany) for discussions on lithium-6 enrichment technology, and Pavan Teki (Aix-Marseille University) for exploratory studies on this topic.

\textit{This work has been carried out within the framework of the EUROfusion Consortium, funded by the European Union via the Euratom Research and Training Programme (Grant Agreement No 101052200 - EUROfusion). Views and opinions expressed are however those of the author(s) only and do not necessarily reflect those of the European Union or the European Commission. Neither the European Union nor the European Commission can be held responsible for them.}

\textit{Richard Pearson is also a co-founder and employee at Kyoto Fusioneering, a Japanese fusion energy company focused on developing technologies including tritium breeding blankets.}

\scriptsize 
\bibliographystyle{unsrt}
\bibliography{lithiumpaper}




\end{document}